\documentclass{PoS}

\title{SU(3)-breaking corrections to the hyperon vector coupling $f_1(0)$ in covariant baryon
chiral perturbation theory}

\ShortTitle{SU(3)-breaking corrections to $f_1(0)$ in Covariant
ChPT}

\author{\speaker{L. S. Geng}\\
        Departamento de F\'{\i}sica Te\'orica and IFIC, Universidad de
Valencia-CSIC, E-46071 Valencia, Spain\\
        E-mail: \email{lsgeng@ific.uv.es}}

\author{J. Martin Camalich\\
        Departamento de F\'{\i}sica Te\'orica and IFIC, Universidad de
Valencia-CSIC, E-46071 Valencia, Spain\\
        E-mail: \email{camalich@ific.uv.es}}

\author{M.J. Vicente Vacas\\
        Departamento de F\'{\i}sica Te\'orica and IFIC, Universidad de
Valencia-CSIC, E-46071 Valencia, Spain\\
        E-mail: \email{vicente@ific.uv.es}}

\abstract{ We report on a recent study of the SU(3)-breaking
corrections to the hyperon vector coupling $f_1(0)$ up to
$\mathcal{O}(p^4)$ in covariant baryon chiral perturbation theory
with dynamical octet and decuplet contributions. The decuplet
contributions are taken into account for the first time in a
covariant ChPT study and are found of similar or even larger size
than the octet ones. We predict positive
SU(3)-breaking corrections to all the four independent $f_1(0)$'s
(assuming isospin symmetry), which are consistent, within
uncertainties, with the latest results from large $N_c$ fits, chiral
quark models, and quenched lattice QCD calculations. We also discuss
briefly the implications of our results for the extraction of 
$V_{us}$ from hyperon decay data.}

\FullConference{6th International Workshop on Chiral Dynamics, CD09\\
        July 6-10, 2009\\
        Bern, Switzerland}

\begin{document}

\section{Introduction}
The Cabibbo-Kobayashi-Maskawa (CKM) matrix ~\cite{Cabibbo:1963yz,Kobayashi:1973fv}
\begin{equation}
V_{CKM}=\left(\begin{array}{ccc}
               V_{ud}&V_{us}&V_{ub}\\
               V_{cd}&V_{cs}&V_{cb}\\
               V_{td}&V_{ts}&V_{tb}
               \end{array}\right)
\end{equation}
plays a very important role in our study and understanding of flavor
physics. Particularly, an accurate value of $V_{us}$ is crucial in
determinations of the other parameters  and in tests of CKM
unitarity. For instance, to test the first row unitarity,
\begin{equation}
|V_{ud}|^2+|V_{us}|^2+|V_{ub}|^2=1,
\end{equation}
one needs to know the values of $V_{ud}$, $V_{us}$, and $V_{ub}$.
Among them, $V_{ub}$ is quite small and can be
neglected at the present precision; $V_{ud}$ can be obtained from
superallowed nuclear beta decays, neutron and pion decays; while
$V_{us}$ can be obtained from kaon decays, hyperon decays, and tau
decays (for a recent review, see Ref.~\cite{Amsler:2008zzb}). In this work, we focus on an important quantity in order to
obtain $V_{us}$ from hyperon decay data -- the $f_1(0)$.

To extract $V_{us}$ from hyperon decay data, one must know the
hyperon vector coupling $f_1(0)$, since experimentally only the product of $|V_{us}
f_1(0)|$ is accessible. Theoretically, $f_1(0)$ is known up to SU(3)
breaking corrections due to the hypothesis of Conservation of Vector
Current (CVC) . To obtain an accurate $f_1(0)$, one then needs to
know the size of SU(3) breaking, which, naively, could be as large
as $30\%$. On the other hand, the Ademollo-Gatto (AG) theorem~\cite{Ademollo:1964sr}
states that
\begin{equation}\label{eq:ag}
f_1(0)=g_v+\mathcal{O}((m_s-m)^2)
\end{equation}
where $m_s$ is the strange quark mass and $m$ is the mass of the
light quarks. The values of $g_V$ are $-\sqrt{\frac{3}{2}}$,
$-\frac{1}{\sqrt{2}}$, $-1$,  $\sqrt{\frac{3}{2}}$,
$\frac{1}{\sqrt{2}}$,  $1$ for  $\Lambda\rightarrow p$,
$\Sigma^0\rightarrow p$,
 $\Sigma^-\rightarrow n$,  $\Xi^-\rightarrow \Lambda$, $\Xi^-\rightarrow \Sigma^0$, and
$\Xi^0\rightarrow\Sigma^+$, respectively. In the isospin-symmetric
limit only four channels, which we take as $\Lambda\rightarrow N$,
$\Sigma\rightarrow N$, $\Xi\rightarrow\Lambda$, and
$\Xi\rightarrow\Sigma$, provide independent information. The AG
theorem not only tells that SU(3)-breaking corrections are of the
order of $10\%$ but also has a very important consequence for a calculation of
$f_1(0)$ in chiral
perturbation theory as we will explain below.

 Theoretical
estimates of SU(3)-breaking corrections to $f_1(0)$ have been
performed in various frameworks, including quark
models~\cite{Donoghue:1986th,Schlumpf:1994fb,Faessler:2008ix},
 large-$N_c$ fits~\cite{FloresMendieta:2004sk}, and
chiral perturbation theory
(ChPT)~\cite{Krause:1990xc,Anderson:1993as,Kaiser:2001yc,Villadoro:2006nj,Lacour:2007wm}.
These SU(3)-breaking corrections have also been studied recently in
quenched lattice QCD (LQCD) calculations  for the two channels:
$\Sigma^-\rightarrow n$~\cite{Guadagnoli:2006gj} and
$\Xi^0\rightarrow\Sigma^+$~\cite{Sasaki:2008ha}.

Compared to earlier ChPT studies~\cite{Krause:1990xc,Anderson:1993as,Kaiser:2001yc,Villadoro:2006nj,Lacour:2007wm}, our work~\cite{Geng:2009ik} contains two improvements:
 \begin{enumerate}
 \item
 We have performed a calculation that fully conserves analyticity and
relativity without introducing any power-counting-restoration (PCR) dependence.
This is possible because the Ademollo-Gatto theorem
tells that up to $\mathcal{O}(p^4)$ no unknown LEC's contribute
and, therefore, no power-counting-breaking terms shows up.
Consequently, up to this order there is no need to apply any
PCR procedure.

\item We have taken into account the contributions of virtual decuplet
baryons.
    They are important because $m_D-m_B\approx0.231$ GeV is similar to the pion mass and
   much smaller than the kaon (eta) mass, where $m_D$ and $m_B$ are the averages
   of the octet- and decuplet-baryon masses, respectively. Therefore, in SU(3) baryon ChPT, one has
to be cautious about the exclusion
   of virtual decuplet baryons.
    As we will show, the decuplet baryons do provide sizable contributions that
   completely change the results obtained with only virtual octet baryons

\end{enumerate}

\section{Formalism}

The baryon vector form factors as probed by the charged $\Delta$S=1
weak current $V^\mu=V_{us}\bar{u}\gamma^\mu s$ are defined by
\begin{equation}
\langle B'\vert V^\mu\vert B\rangle
=V_{us}\bar{u}(p')\left[\gamma^\mu f_1(q^2)+\frac{2i
\sigma^{\mu\nu}q_\nu}{M_{B'}+M_B}f_2(q^2)+\frac{2 q^\mu
}{M_{B'}+M_B}f_3(q^2)\right]u(p),
\end{equation}
where $q=p'-p$.   We will parameterize the SU(3)-breaking
corrections order-by-order in the covariant chiral expansion as
follows:
\begin{equation}
f_1(0)=g_V\left( 1+\delta^{(2)}+\delta^{(3)}+\cdots\right)
\label{eq:Adem-Gatt},
\end{equation}
where $\delta^{(2)}$ and $\delta^{(3)}$ are the leading and
next-to-leading order SU(3)-breaking corrections induced by loops,
corresponding to $\mathcal{O}(p^3)$ and $\mathcal{O}(p^4)$ chiral
calculations.

The calculation of the virtual octet-baryon contributions is
standard and details can be found in Ref.~\cite{Geng:2009ik}. Here we would like to
stress the calculation of the virtual decuplet-baryon contributions.
A fully consistent and problem-free description of spin-3/2
particles in a quantum-field-theory framework is not yet possible,
although progress has been made in the past few decades (see e.g. Refs.~\cite{Pascalutsa:2000kd,Hacker:2005fh}). In the framework of
effective field theories, such as ChPT, a solution has been proposed
in Ref.~\cite{Pascalutsa:2000kd}, where the couplings of spin-3/2
particles to spin-1/2 particles satisfy spin-3/2 gauge symmetry and are referred to ``consistent''
couplings. In addition to
this formal appealing nature, certain ChPT calculations performed with these
consistent couplings have shown better convergence behavior than the
same calculations done in the ``conventional coupling'' scheme (see e.g. Refs.~\cite{Geng:2009hh}). In the present work, we have adopted the
``consistent'' couplings to describe the interactions of
decuplet-baryons with octet-baryons. Details can
be found in Ref.~\cite{Geng:2009ik}.
\section{Results and discussions}
\subsection{SU(3)-breaking corrections to $f_1(0)$ due to octet contributions up to $\mathcal{O}(p^4)$}

\begin{figure}[t]

\centerline{\includegraphics[scale=0.65,angle=270]{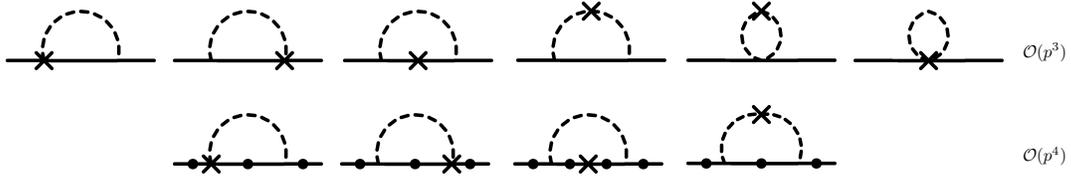}}
\caption{Feynman diagrams contributing to the SU(3)-breaking
corrections to the hyperon vector coupling $f_1(0)$ up to
$\mathcal{O}(p^4)$. The solid lines correspond to baryons and dashed
lines to mesons; crosses indicate the coupling of the external
current; black dots denote mass splitting insertions. We have not
shown explicitly those diagrams corresponding to wave function
renormalization, which have been taken into account in the
calculation. \label{fig:doctet}}
\end{figure}
All the diagrams contributing to $f_1(0)$ up to $\mathcal{O}(p^4)$
are shown in Fig.~\ref{fig:doctet}, where the leading and
next-to-leading order SU(3)-breaking corrections are given by the
diagrams in the first and second row, respectively.

The $\mathcal{O}(p^3)$ results are quite compact and have the
following structure for the transition $i\rightarrow j$:
\begin{eqnarray}\label{eq:sumo3}
 \delta_B^{(2)}(i\rightarrow j)&=&\sum_{M=\pi,\eta,K}\beta^\mathrm{BP}_M H_\mathrm{BP}(m_M)+\sum_{M=\pi,\eta} \beta^\mathrm{MP}_{M} H_\mathrm{MP}(m_M,m_K)+\sum_{M=\pi,\eta,K} \beta^\mathrm{KR}_M H_\mathrm{KR}(m_M)\nonumber\\
&&-\frac{3}{8}\sum_{M=\pi,\eta}H_\mathrm{TD1}(m_M,m_K)+\frac{3}{8}\sum_{M=\pi,\eta}H_\mathrm{TD2}(m_M)
+\frac{3}{4}H_\mathrm{TD2}(m_K)\nonumber\\
&&+\frac{1}{2}\sum_{M=\pi,\eta,K}(\beta^\mathrm{WF}_M(i)+\beta^\mathrm{WF}_M(j))H_\mathrm{WF}(m_M),
\end{eqnarray}
where $\beta^\mathrm{BP}$, $\beta^\mathrm{MP}$, $\beta^\mathrm{KR}$,
and $\beta^\mathrm{WF}$ , and the loop functions $H_\mathrm{BP}$,
$H_\mathrm{MP}$, $H_\mathrm{KR}$, $H_\mathrm{TD1}$,
$H_\mathrm{TD2}$, and $ H_\mathrm{WF}$ are given in Appendix A of
Ref.~\cite{Geng:2009ik}. It is interesting to point out that although separately
these loop functions are divergent (scale-dependent) and some of
them contain power-counting-breaking pieces ($H_\mathrm{KR}$ and
$H_\mathrm{MP}$), the overall contributions are finite and do not
break power-counting. This is an explicit manifestation of the
AG theorem.

 Similar to the IRChPT study of
Ref.~\cite{Lacour:2007wm}, the $\mathcal{O}(p^4)$ results contain
higher-order divergences. We have removed the infinities using the
modified minimal-subtraction ($\overline{MS}$) scheme. 
The remaining scale dependence is shown in
Fig.~2 of Ref.~\cite{Geng:2009ik}, which is rather mild for most cases except for the
$\Sigma\rightarrow N$ transition.  The scale dependence can also be used
to estimate the size of higher-order contributions by varying $\mu$ in a
reasonable range. In the following, we present the results by
varying $\mu$ from 0.7 to 1.3 GeV. It should be mentioned that if we
had adopted the same method as Ref.~\cite{Villadoro:2006nj} to
calculate the $\mathcal{O}(p^4)$ contributions, i.e., by expanding
the results and keeping only those linear in baryon mass splittings,
our $\mathcal{O}(p^4)$ results would have been convergent.

\begin{table}[htbp]
      \renewcommand{\arraystretch}{1.5}
     \setlength{\tabcolsep}{0.4cm}
     \begin{center}
     \caption{Values for the masses and couplings appearing in the calculation of
the SU(3)-breaking corrections to $f_1(0)$. \label{table:para}}
\begin{tabular}{cl|cl}
\hline\hline
 $D$& $0.8$ & $M_B$ & $1.151$ GeV \\
 $F$& $0.46$ & $M_D$ &$1.382$ GeV\\
 $f_\pi$ & $0.0924$ GeV & $M_0$ & $1.197$ GeV \\
 $F_0$& $1.17f_\pi$ & $b_D$ & $-0.0661$ GeV$^{-1}$\\
 $m_\pi$ & $0.138$  GeV&$b_F$ & $0.2087$ GeV$^{-1}$\\
 $m_K$ & $0.496$  GeV & $M_{D0}$ & $1.216$ GeV \\
 $m_\eta$ & $0.548$ GeV & $\gamma_M$ & $0.3236$ GeV$^{-1}$\\
 $\mathcal{C}$& 1.0\\
\hline\hline
\end{tabular}
\end{center}
\end{table}
Table \ref{table:octetfull} shows the SU(3)-breaking corrections in
the notation of Eq.~(\ref{eq:Adem-Gatt}). For the sake of comparison, we also
list the numbers obtained in HBChPT~\cite{Villadoro:2006nj} and
IRChPT~\cite{Lacour:2007wm}. The numerical values are obtained with
the parameters given in Table \ref{table:para}.  As in
Ref.~\cite{Geng:2008mf} we have used an average $F_0=1.17 f_\pi$
with $f_\pi=92.4$ MeV. It should be pointed out that the HBChPT and
the IRChPT results are obtained using $f_\pi$.

First, we note that in three of the four cases, the $\delta^{(3)}$
numbers are smaller than the $\delta^{(2)}$ ones. The situation is
similar in IRChPT  but quite different in HBChPT. In the HBChPT
calculation~\cite{Villadoro:2006nj}, the $\delta^{(3)}$ contribution
is larger than the $\delta^{(2)}$ one for all the four
transitions.\footnote{What we denote by $\delta^{(3)}$  is the sum of
those labeled by $\alpha^{(3)}$ and $\alpha^{(1/M)}$ in
Ref.~\cite{Villadoro:2006nj}.}  On
the other hand, the results of the present work and those of
IRChPT~\cite{Lacour:2007wm}, including the contributions of
different chiral orders,
 are qualitatively similar. They are both very different from the HBChPT predictions, even for the signs in three of the
four cases. Obviously, as stressed in Ref.~\cite{Lacour:2007wm}, one
should trust more the covariant  than the HB
results, which have to be treated with caution whenever $1/M$ recoil
corrections become large, as in the present case~\cite{Villadoro:2006nj}.

\begin{table}[t]
      \renewcommand{\arraystretch}{1.5}
     \setlength{\tabcolsep}{0.4cm}
     \caption{Octet contributions to the SU(3)-breaking corrections to $f_1(0)$ (in percentage). The central values of the  $\mathcal{O}(p^4)$ results are calculated with $\mu=1$ GeV and the uncertainties are
obtained by varying $\mu$ from 0.7 to 1.3
GeV.\label{table:octetfull}}
\begin{center}
\begin{tabular}{c|ccc|c|c}
\hline\hline &\multicolumn{3}{c|} {present work}&
HBChPT~\cite{Villadoro:2006nj} & IRChPT~\cite{Lacour:2007wm}\\\hline
                        & $\delta^{(2)}$&  $\delta^{(3)}$ & $\delta^{(2)}+\delta^{(3)}$& $\delta^{(2)}+\delta^{(3)}$& $\delta^{(2)}+\delta^{(3)}$\\
            \hline
 $\Lambda\rightarrow N$ & $-3.8$  &  $0.2^{+1.2}_{-0.9}$ & $-3.6^{+1.2}_{-0.9}$  & $2.7$ & $-5.7\pm2.1$ \\
 $\Sigma\rightarrow N$  & $-0.8$ &  $4.7^{+3.8}_{-2.8}$ & $3.9^{+3.8}_{-2.8}$   & $4.1$ & $2.8\pm0.2$\\
 $\Xi\rightarrow\Lambda$& $-2.9$  &  $1.7^{+2.4}_{-1.8}$ & $-1.2^{+2.4}_{-1.8}$ & $4.3$ & $-1.1\pm1.7$\\
 $\Xi\rightarrow\Sigma$ & $-3.7$  &  $-1.3^{+0.3}_{-0.2}$ & $-5.0^{+0.3}_{-0.2}$  & $0.9$ & $-5.6\pm1.6$\\
 \hline\hline
\end{tabular}
\end{center}
\end{table}

\begin{figure}[t]
\centerline{\includegraphics[scale=0.65,angle=270]{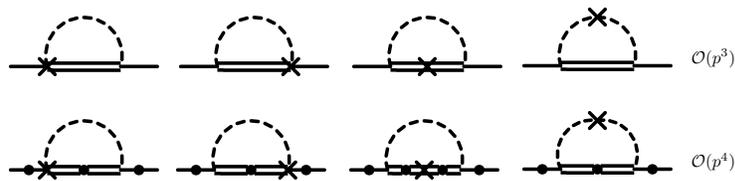}}
\caption[]{\label{fig:ddecup} Feynman diagrams contributing to the leading and
next-to-leading order SU(3)-breaking corrections to the hyperon
vector coupling $f_1(0)$, through dynamical decuplet baryons. The
notations are the same as those of Fig.~\ref{fig:doctet} except that
double lines indicate decuplet baryons. We have not shown explicitly
those diagrams corresponding to wave function renormalization, which
have been included in the calculation. }
\end{figure}

\subsection{SU(3)-breaking corrections to $f_1(0)$ induced by dynamical decuplet baryons up to $\mathcal{O}(p^4)$}
Fig.~\ref{fig:ddecup} shows the diagrams that contribute to
SU(3)-breaking corrections to  $f_1(0)$ with dynamical decuplet
baryons up to $\mathcal{O}(p^4)$.  It should be noted that unlike in
the HBChPT case~\cite{Villadoro:2006nj}, Kroll-Rudermann (KR) kind
of diagrams also contribute. In fact, using the consistent coupling
scheme of Ref.~\cite{Pascalutsa:2000kd}, there are four KR diagrams:
Two are from minimal substitution in the derivative of the
pseudoscalar fields and
 the other
two are from minimal substitution in the derivative of the decuplet
fields.

As in the previous case, the $\mathcal{O}(p^4)$ results contain again
higher-order divergences, which have been removed by the $\overline{MS}$
procedure  with the remaining scale dependence shown in Fig.~4
of Ref.~\cite{Geng:2009ik}.  In this case, unlike in the previous
case, the divergences cannot be removed by expanding and keeping
only terms linear in baryon and decuplet mass splittings.

The numerical results obtained with the parameter values given in
Table \ref{table:para} are summarized in Table
\ref{table:decupfullpas}. It can be seen that at $\mathcal{O}(p^3)$,
the decuplet contributions are relatively small compared to the
octet ones at the same order. On the other hand, the
$\mathcal{O}(p^4)$ contributions are sizable and all of them have
positive signs.

\begin{table}[t]
      \renewcommand{\arraystretch}{1.5}
     \setlength{\tabcolsep}{0.4cm}
     \begin{center}
     \caption{Decuplet contributions to the SU(3)-breaking corrections to $f_1(0)$ (in percentage). The central values of the $\mathcal{O}(p^4)$ result are calculated with $\mu=1$ GeV and the uncertainties are obtained by varying $\mu$ from 0.7 to 1.3 GeV.
     \label{table:decupfullpas}}
\begin{tabular}{c|ccc|ccc}
\hline\hline &\multicolumn{3}{c}{Present
work}&\multicolumn{3}{c}{HBChPT}\\\hline
& $\delta^{(2)}$&  $\delta^{(3)}$ & $\delta^{(2)}+\delta^{(3)}$  & $\delta^{(2)}$&  $\delta^{(3)}$ & $\delta^{(2)}+\delta^{(3)}$\\
            \hline
 $\Lambda\rightarrow N$ &    $0.7$ & $3.0^{+0.1}_{-0.1}$ & $3.7^{+0.1}_{-0.1}$ & $1.8$ & $1.3$ & $3.1$ \\
 $\Sigma\rightarrow N$  & $-1.4$ & $6.2^{+0.4}_{-0.3}$ & $4.8^{+0.4}_{-0.3}$& $-3.6$ & $8.8$ & $5.2$\\
 $\Xi\rightarrow\Lambda$&  $-0.02$ & $5.2^{+0.4}_{-0.3}$ & $5.2^{+0.4}_{-0.3}$ &$-0.05$ & $4.2$ & $4.1$ \\
 $\Xi\rightarrow\Sigma$ &  $0.7$ & $6.0^{+1.9}_{-1.4}$ & $6.7^{+1.9}_{-1.4}$ &$1.9$ & $-0.2$ & $1.7$\\
 \hline\hline
\end{tabular}
\end{center}
\end{table}

In Table \ref{table:decupfullpas}, the numbers denoted by HBChPT
are obtained by taking our covariant results to the heavy-baryon limit, as
explained in detail in Ref.~\cite{Geng:2009ik}. They 
are different from those of Ref.~\cite{Villadoro:2006nj} and the origin
of the difference is also discussed in Ref.~\cite{Geng:2009ik}.

\subsection{Full results and comparison with other approaches}

\begin{table}[t]
      \renewcommand{\arraystretch}{1.5}
     \setlength{\tabcolsep}{0.4cm}
     \begin{center}
     \caption{SU(3)-breaking corrections to $f_1(0)$ up to $\mathcal{O}(p^4)$ (in percentage), including both
the octet and the decuplet contributions.
     \label{table:full}}
\begin{tabular}{c|ccc}
\hline\hline
& $\delta^{(2)}$&  $\delta^{(3)}$ & $\delta^{(2)}+\delta^{(3)}$ \\
            \hline
 $\Lambda\rightarrow N$ &    $-3.1$ & $3.2^{+1.3}_{-1.0}$ & $0.1^{+1.3}_{-1.0}$ \\
 $\Sigma\rightarrow N$  & $-2.2$ & $10.9^{+4.2}_{-3.1}$ & $8.7^{+4.2}_{-3.1}$\\
 $\Xi\rightarrow\Lambda$&  $-2.9$ & $6.9^{+2.8}_{-2.1}$ & $4.0^{+2.8}_{-2.1}$\\
 $\Xi\rightarrow\Sigma$ &  $-3.0$ & $4.7^{+2.2}_{-1.6}$ & $1.7^{+2.2}_{-1.6}$ \\
 \hline\hline
\end{tabular}
\end{center}
\end{table}
Summing the octet  and the decuplet contributions, we obtain the
numbers shown in Table \ref{table:full}. Two things are noteworthy.
First, the convergence is slow, even taking into account the scale
dependence of the $\delta^{(3)}$ corrections. Second, for three of
the four transitions, the $\delta^{(3)}$ corrections have a
different sign than the $\delta^{(2)}$ ones.

In Table \ref{table:comparison}, we compare our results with those
obtained from other approaches, including large $N_c$
fits~\cite{FloresMendieta:2004sk}, quark
models~\cite{Donoghue:1986th,Schlumpf:1994fb, Faessler:2008ix}, and
two quenched LQCD
calculations~\cite{Guadagnoli:2006gj,Sasaki:2008ha}. The large $N_c$
results in general favor positive corrections, which are consistent
with our central values. Two of the quark models predict negative
corrections, while that of Ref.~\cite{Faessler:2008ix} favors
positive corrections. It is interesting to note that in
Ref.~\cite{Faessler:2008ix} the valence quark effects give negative
contributions, as in the other two quark models; on the other hand, the chiral
effects provide positive contributions, resulting in net positive
corrections. Our numbers also agree, within uncertainties, with the
quenched LQCD ones. 
\begin{table}[t]
      \renewcommand{\arraystretch}{1.5}
     \setlength{\tabcolsep}{0.2cm}
     \caption{SU(3)-breaking corrections (in percentage) to $f_1(0)$ obtained
in different approaches.
     \label{table:comparison}}
\begin{center}
\begin{tabular}{c|c|c|ccc|c}
\hline\hline & Present work & Large $N_c$ &
\multicolumn{3}{c|}{Quark model} & Quenched LQCD \\\hline &
&Ref.~~\cite{FloresMendieta:2004sk}   & Ref.~\cite{Donoghue:1986th}
&Ref.~\cite{Schlumpf:1994fb} & Ref.~\cite{Faessler:2008ix} &
\\\hline
 $\Lambda\rightarrow N$ &    $0.1^{+1.3}_{-1.0}$ & $2\pm2$ & $-1.3$ & $-2.4$ & $0.1$ & \\
 $\Sigma\rightarrow N$  & $8.7^{+4.2}_{-3.1}$ &  $4\pm3$ &  $-1.3$ & $-2.4$ & $0.9$ & $-1.2\pm2.9\pm4.0$~\cite{Guadagnoli:2006gj}\\
 $\Xi\rightarrow\Lambda$ &  $4.0^{+2.8}_{-2.1}$&  $4\pm4$ &  $-1.3$ & $-2.4$ & $2.2$ & \\
 $\Xi\rightarrow\Sigma$ &  $1.7^{+2.2}_{-1.6}$ &  $8\pm5$ &  $-1.3$ & $-2.4$ & $4.2$ & $-1.3\pm1.9$~\cite{Sasaki:2008ha}\\
 \hline\hline
\end{tabular}
\end{center}
\end{table}

\subsection{Implications for the extraction of $V_{us}$ from hyperon decay data}

In the following, we  briefly discuss the implications of our results for
the extracting of $V_{us}$ from hyperon decay data. There have been several previous
attempts to extract $V_{us}$ using hyperon semileptonic decay data
\cite{Cabibbo:2003cu,Cabibbo:2003ea,FloresMendieta:2004sk,Mateu:2005wi}.
As discussed in Ref.~\cite{Mateu:2005wi} a rather clean
determination of $f_1 V_{us}$ can be done by using $g_1/f_1$ and the
decay rates from experiment and taking for $g_2$ and $f_2$ their
SU(3) values. This latter approximation is supported by the fact
that their contributions to the decay rate are reduced by kinematic
factors (See, for instance, Eq. (10) of
Ref.~\cite{FloresMendieta:2004sk}). Using the values of $f_1 V_{us}$
compiled in Table 3 of Ref.~\cite{Mateu:2005wi} and our results for
$f_1$ we get
\begin{equation}
\label{eq:vus} V_{us} =0.2177\pm 0.0030,
\end{equation}
where the error includes only the experimental errors and the
uncertainties related to the scale dependence. This value is lower
than the value obtained in
Refs.~\cite{Cabibbo:2003cu,Cabibbo:2003ea}, which  is easy to understand
because our procedure is similar to that of Refs.~\cite{Cabibbo:2003cu,Cabibbo:2003ea}
but our calculated SU(3) breaking corrections are positive while they are assumed to
be zero in Refs.~\cite{Cabibbo:2003cu,Cabibbo:2003ea}. 

\section{Summary and conclusions}
The CKM matrix element $V_{us}$ plays a very important role in
studies of flavor physics. At present, its most accurate value
is obtained from kaon decays. However, it will be of vital importance
to be able to extract its value also from other sources, such as hyperon decay data.

To extract $V_{us}$ from hyperon decay data, one must know very accurately the
value of $f_1(0)$ since experimentally only the product of $V_{us} f_1(0)$ is accessible.
Chiral perturbation theory provides a model-independent prediction for
$f_1(0)$ up to $\mathcal{O}(p^4)$, thanks to the Ademollo-Gatto theorem.

In this work, we have performed a fully-covariant calculation up to $\mathcal{O}(p^4)$ 
in chiral perturbation theory taking 
into account the contributions of virtual decuplet baryons, which are found to be
comparable in size  to the virtual octet-baryon contributions. Our study predicts
positive SU(3) breaking corrections to $f_1(0)$ in all the four channels,  $\Lambda\rightarrow N$,
$\Sigma\rightarrow N$, $\Xi\rightarrow\Lambda$, and
$\Xi\rightarrow\Sigma$,  in agreement with those obtained from the large $N_c$ fits. 
 We encourage the use of our calculated $f_1(0)$ in future analysis of
hyperon decay data.

\section{Acknowledgments}
This work was partially supported by the  MEC grant  FIS2006-03438 and the European Community-Research Infrastructure
Integrating Activity Study of Strongly Interacting Matter (Hadron-Physics2, Grant Agreement 227431) under the Seventh Framework Programme of EU. L.S.G. acknowledges support from the MICINN in the Program 
``Juan de la Cierva.'' J.M.C. acknowledges the same institution for a FPU grant.


\begin{thebibliography}{99}


\bibitem{Cabibbo:1963yz}
  N.~Cabibbo,
  Phys.\ Rev.\ Lett.\  {\bf 10}, 531 (1963).

\bibitem{Kobayashi:1973fv}
  M.~Kobayashi and T.~Maskawa,
  Prog.\ Theor.\ Phys.\  {\bf 49}, 652 (1973).


\bibitem{Amsler:2008zzb}
  C.~Amsler {\it et al.}  [Particle Data Group],
  Phys.\ Lett.\  B {\bf 667}, 1 (2008).


\bibitem{Ademollo:1964sr}
  M.~Ademollo and R.~Gatto,
  Phys.\ Rev.\ Lett.\  {\bf 13}, 264 (1964).

\bibitem{Donoghue:1986th}
  J.~F.~Donoghue, B.~R.~Holstein and S.~W.~Klimt,
  Phys.\ Rev.\  D {\bf 35}, 934 (1987).

\bibitem{Schlumpf:1994fb}
  F.~Schlumpf,
  Phys.\ Rev.\  D {\bf 51}, 2262 (1995).

\bibitem{Faessler:2008ix}
  A.~Faessler, T.~Gutsche, B.~R.~Holstein, M.~A.~Ivanov, J.~G.~Korner and V.~E.~Lyubovitskij,
  Phys.\ Rev.\  D {\bf 78}, 094005 (2008).

\bibitem{FloresMendieta:2004sk}
  R.~Flores-Mendieta,
  Phys.\ Rev.\  D {\bf 70}, 114036 (2004).
\bibitem{Krause:1990xc}
  A.~Krause,
  Helv.\ Phys.\ Acta {\bf 63}, 3 (1990).

\bibitem{Anderson:1993as}
  J.~Anderson and M.~A.~Luty,
  Phys.\ Rev.\  D {\bf 47}, 4975 (1993).

\bibitem{Kaiser:2001yc}
  N.~Kaiser,
  Phys.\ Rev.\  C {\bf 64}, 028201 (2001).

\bibitem{Villadoro:2006nj}
  G.~Villadoro,
  Phys.\ Rev.\  D {\bf 74}, 014018 (2006).


\bibitem{Lacour:2007wm}
  A.~Lacour, B.~Kubis and U.~G.~Meissner,
  JHEP {\bf 0710}, 083 (2007).


\bibitem{Guadagnoli:2006gj}
  D.~Guadagnoli, V.~Lubicz, M.~Papinutto and S.~Simula,
  Nucl.\ Phys.\  B {\bf 761}, 63 (2007).


\bibitem{Sasaki:2008ha}
  S.~Sasaki and T.~Yamazaki,
  Phys.\ Rev.\  D {\bf 79}, 074508 (2009).



\bibitem{Geng:2009ik}
  L.~S.~Geng, J.~Martin Camalich and M.~J.~Vicente Vacas,
  Phys.\ Rev.\  D {\bf 79}, 094022 (2009).

\bibitem{Pascalutsa:2000kd}
  V.~Pascalutsa,
  Phys.\ Lett.\  B {\bf 503}, 85 (2001).








\bibitem{Hacker:2005fh}
  C.~Hacker, N.~Wies, J.~Gegelia and S.~Scherer,
  Phys.\ Rev.\  C {\bf 72}, 055203 (2005).




\bibitem{Geng:2009hh}
  L.~S.~Geng, J.~Martin Camalich and M.~J.~Vicente Vacas,
  Phys.\ Lett.\  B {\bf 676}, 63 (2009).


\bibitem{Geng:2008mf}
  L.~S.~Geng, J.~M.~Camalich, L.~Alvarez-Ruso and M.~J.~V.~Vacas,
  Phys.\ Rev.\ Lett.\  {\bf 101}, 222002 (2008).


\bibitem{Cabibbo:2003cu}
  N.~Cabibbo, E.~C.~Swallow and R.~Winston,
  Ann.\ Rev.\ Nucl.\ Part.\ Sci.\  {\bf 53}, 39 (2003).

\bibitem{Cabibbo:2003ea}
  N.~Cabibbo, E.~C.~Swallow and R.~Winston,
  Phys.\ Rev.\ Lett.\  {\bf 92}, 251803 (2004).



\bibitem{Mateu:2005wi}
  V.~Mateu and A.~Pich,
  JHEP {\bf 0510}, 041 (2005).






 















\end{thebibliography}
\end{document}